\documentclass{emulateapj}

\usepackage{graphicx}
\usepackage{epsfig}
\usepackage{txfonts}
\usepackage{longtable}
\usepackage{lscape}
\usepackage{natbib}

\shorttitle{The scattering polarization of Ly$\alpha$}
\shortauthors{\v{S}t\v{e}p\'an et al.}

\begin{document}

\title{The Hanle effect of Ly$\alpha$ in an MHD model of the Solar Transition Region}

\author{J. \v{S}t\v{e}p\'an\altaffilmark{1,2}, 
J. Trujillo Bueno\altaffilmark{2,3,4}, 
M. Carlsson\altaffilmark{5,6},
and J. Leenaarts\altaffilmark{5,7}}
\altaffiltext{1}{Astronomical Institute ASCR, Fri\v{c}ova 298, 251\,65 Ond\v{r}ejov, Czech Republic}
\altaffiltext{2}{Instituto de Astrof\'isica de Canarias, E-38205 La Laguna, 
Tenerife, Spain}
\altaffiltext{3}{Departamento de Astrof\'isica, Facultad de F\'isica, 
Universidad de La Laguna, Tenerife, Spain}
\altaffiltext{4}{Consejo Superior de Investigaciones Cient\'ificas, Spain}
\altaffiltext{5}{Institute of Theoretical Astrophysics, University of Oslo, P.O. Box 1029 Blindern, N-0315 Oslo, Norway}
\altaffiltext{6}{Center of Mathematics for Applications, University of Oslo, P.O. Box 1053 Blindern, N-0315 Oslo, Norway}
\altaffiltext{7}{Utrecht University, P.O. Box 80\,000 NL--3508 TA Utrecht, The Netherlands}

\begin{abstract}
In order to understand the heating of the solar corona it is crucial to obtain empirical information on the magnetic field in its lower boundary (the transition region). To this end, we need to measure and model the linear polarization produced by scattering processes in strong UV lines, such as the hydrogen Ly$\alpha$ line. The interpretation of the observed Stokes profiles will require taking into account that the outer solar atmosphere is highly structured and dynamic, and that the height of the transition region may well vary from one place in the atmosphere to another. Here we report on the Ly$\alpha$ scattering polarization signals we have calculated in a realistic model of an enhanced network region, resulting from a state-of-the-art radiation MHD simulation. This model is characterized by spatially complex variations of the physical quantities at transition region heights. The results of our investigation lead us to emphasize that scattering processes in the upper solar chromosphere should indeed produce measurable linear polarization in Ly$\alpha$. More importantly, we show that via the Hanle effect the model's magnetic field produces significant changes in the emergent $Q/I$ and $U/I$ profiles. Therefore, we argue that by measuring the polarization signals produced by scattering processes and the Hanle effect in Ly$\alpha$ and contrasting them with those computed in increasingly realistic atmospheric models, we should be able to decipher the magnetic, thermal and dynamic structure of the upper chromosphere and transition region of the Sun.
\end{abstract}


\section{Introduction\label{sec:intro}}

In order to understand the heating of the solar corona it is crucial to obtain empirical information on the strength and orientation of the magnetic field in its lower boundary (the transition region), where over very short distances (${\sim}100$ km) the temperature suddenly rises from $<10^4$ K to $>10^6$ K and the atmospheric plasma changes from partially to practically fully ionized \citep[e.g.,][]{golub-pasachoff10}. To this end, we need to measure and model observables sensitive to the magnetic field of the solar transition region. 

In a recent paper, \citet{jtb-stepan-casini11} argued that the hydrogen Ly$\alpha$ line is expected to show measurable scattering polarization when observing the solar disk, and that via the Hanle effect the line-center linear polarization amplitude shows a good sensitivity to magnetic field strengths between 10 and 100 G, approximately. These conclusions were reached through detailed radiative transfer calculations in semi-empirical and hydrodynamical  models of the solar atmosphere, assuming complete frequency redistribution (CRD) and neglecting quantum interference between the two upper levels of the Ly$\alpha$ line (see \citet{stepanjtb11a} for details on the atomic model and numerical method of solution). As shown by \citet{belluzzi-trujillobueno-stepan}, these last two approximations are suitable for estimating the polarization signals at the core of Ly$\alpha$, which is the spectral line region where the Hanle effect in Ly$\alpha$ operates.\footnote{Partial frequency redistribution (PRD) and $J$-state interference can, however, produce large linear polarization signals in the wings of the $Q/I$ profile of Ly$\alpha$, especially for close-to-the-limb line of sights \citep[see][]{belluzzi-trujillobueno-stepan}.}

The above-mentioned investigations were carried out using one-dimensional models of the extended solar atmosphere, such as the semi-empirical models of \citet{falc} and the hydrodynamical models of \citet{carlsson-stein-1997}. However, both observations \citep[e.g.,][]{vourlidas-2010} and simulations \citep[e.g.,][]{leenaarts12} show that the outer solar atmosphere is highly structured and dynamic, departing radically from a uniform, plane-parallel configuration. Thus, the height of the transition region must vary from one place in the atmosphere to another, most probably delineating a highly corrugated surface. It is in this type of spatially complex plasmas where strong transition region lines like Ly$\alpha$ originate. The presence of horizontal atmospheric inhomogeneities in a stellar atmosphere breaks the axial symmetry of the radiation field within the medium, and this type of symmetry breaking may produce changes in the scattering polarization signals that could be confused with those caused by the Hanle effect \citep[e.g.,][]{manso-jtb11,anusha-nagendra}. It is therefore important to investigate the problem of scattering polarization and the Hanle effect in Ly$\alpha$ using spatially complex models of the outer solar atmosphere, where the physical quantities may vary abruptly not only along the radial direction but also horizontally. To this end, \citet{stepanjtb12a,stepanjtb12b} have developed a three-dimensional (3D), multilevel radiative transfer code for doing simulations of the spectral line polarization caused by scattering processes and the Hanle and Zeeman effects within the framework of the quantum theory of polarization described in \citet{landi-landolfi04}.

The present paper represents a first step towards that goal. Given the complexity of the radiative transfer problem of the Hanle effect in 3D model atmospheres here we start by considering a two-dimensional (2D) model atmosphere characterized by spatially complex variations of the physical quantities at chromospheric and transition region heights, taken from the 3D radiation magneto-hydrodynamic (MHD) simulations mentioned below. Of particular interest here is to compare the $Q/I$ and $U/I$ profiles computed assuming $B=0$ G at each spatial grid point of the considered MHD atmospheric model with the fractional linear polarization signals calculated taking into account the Hanle effect caused by the model's magnetic field.

\section{Formulation of the radiative transfer problem\label{sec:formulation}}

\begin{figure}[t]
\centering
\includegraphics[width=\columnwidth]{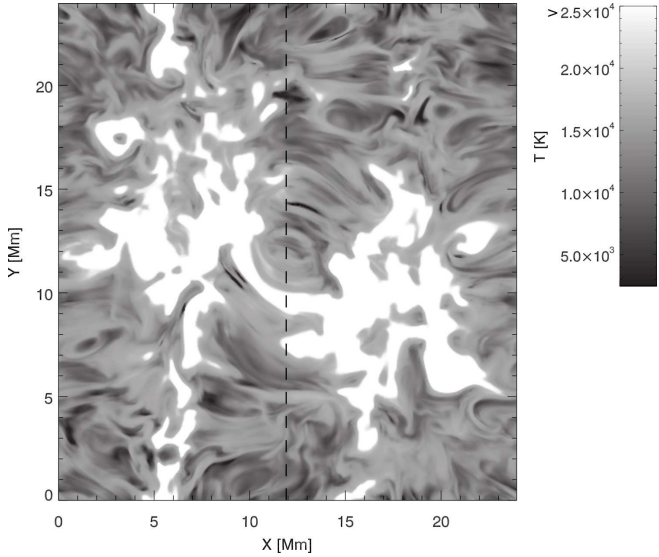}  
\caption[]{The horizontal variation of the kinetic temperature of the 3D model at a height of 2000 km. The dashed line indicates the X-position of the Z-Y slice chosen for obtaining the 2D model studied in this Letter.}
\label{fig:figure-1}
\end{figure}

We have selected a snapshot from a state-of-the-art radiation MHD simulation performed with the Bifrost code \citep{gudiksen11} taking into account non-equilibrium hydrogen ionization. A snapshot from the same simulation (but 30 seconds earlier in time), which has a magnetic field configuration representative of an enhanced network region, was used by \citet{leenaarts12} and we refer the reader to that paper for details on the 3D structure. Figure~\ref{fig:figure-1} illustrates the snapshot we have selected from this simulation. The magnetic field has a predominantly bipolar structure, with magnetic field lines connecting two clusters of photospheric magnetic concentrations of opposite polarity and reaching chromospheric and coronal heights. In this Letter we consider a two-dimensional (2D) model atmosphere extracted from this 3D snapshot; its physical quantities are those found within the Z-Y plane at X=12 Mm (see the dashed line in Fig.~\ref{fig:figure-1}). Since the magnetic field lines in this 3D snapshot are approximately perpendicular to such a Z-Y plane and the temperature between the magnetic patches shows filamentary structure that follows the general orientation of the magnetic field, we believe that the 2D approximation is reasonable for computing the emergent Stokes profiles for line of sights (LOS) with azimuths $\chi=0^{\circ}$ or $\chi=180^{\circ}$ (i.e., for LOS contained in any Z-X plane).

\begin{figure*}[t]
\centering
\includegraphics[width=\textwidth]{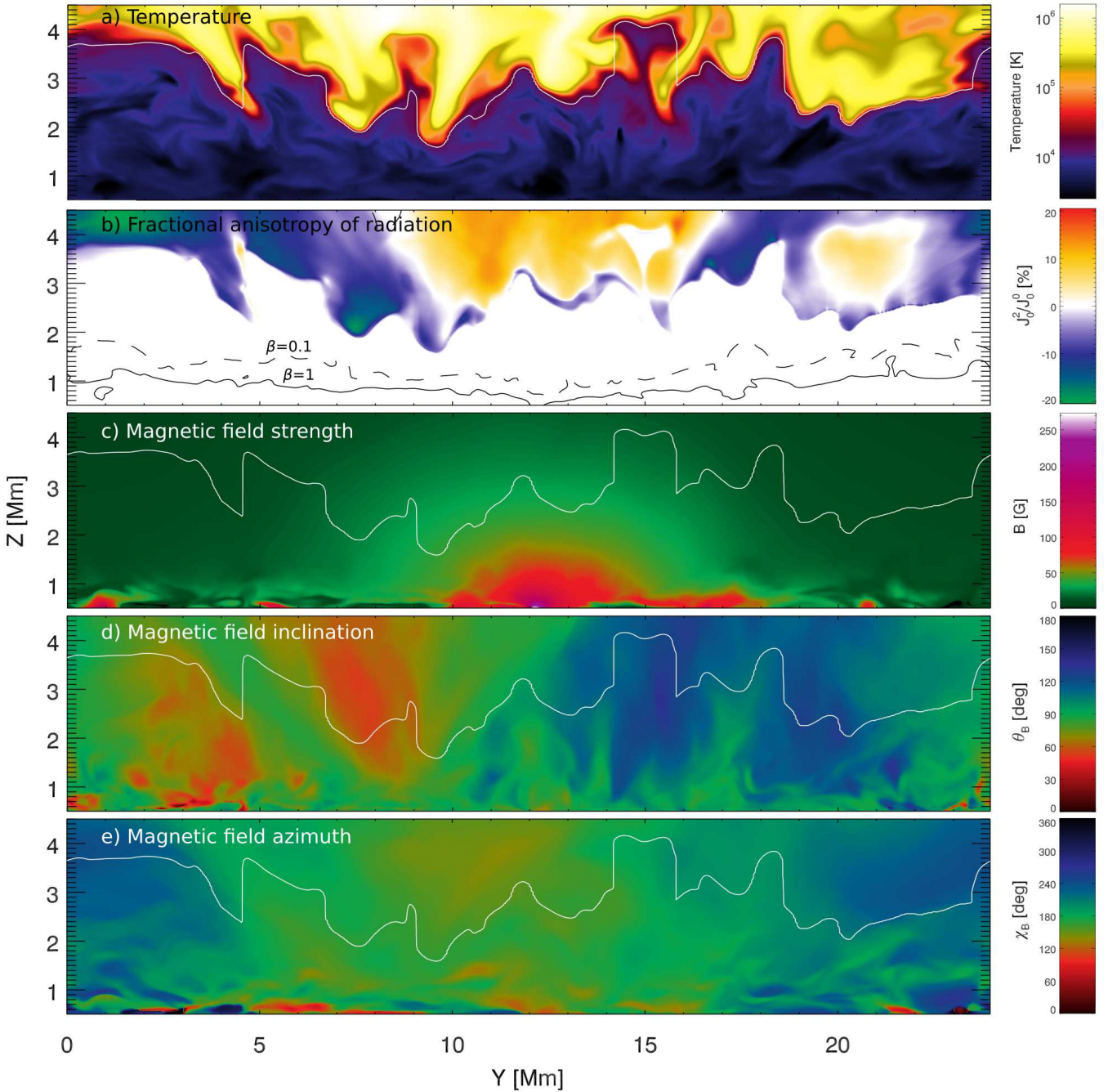}
\caption[]{Physical quantities across the vertical 2D slice defined in Fig.~\ref{fig:figure-1}. From top to bottom: (a) temperature, (b) the fractional anisotropy of the Ly$\alpha$ radiation field (the black contours show the plasma $\beta$ parameter), (c) magnetic field strength $B$, (d) inclination of magnetic field with respect to the positive Z-axis, and (e) azimuth of magnetic field with respect to the positive X-axis. The white line in each panel indicates the height, associated to each horizontal Y-point, where the line-center optical depth is unity along a LOS with $\mu=0.3$, and we point out that it is very similar to that corresponding to other $\mu$ values. Note that in one-dimensional semi-empirical models the transition region to the coronal temperatures occurs instead at a rather precise height (e.g., at about 2200 km in the FAL-C model), which approximately coincides with the height where the Ly$\alpha$ line-center optical depth is unity along the LOS. In both types of atmospheric models (MHD and semi-empirical) the line-core Ly$\alpha$ radiation originates around such transition region heights.}
\label{fig:figure-2}
\end{figure*}

The Ly$\alpha$ line results from two blended transitions between the ground level of hydrogen, $1s\,{}^2\! S_{\!1/2}$, and the $2p\,{}^2\!P_{\!1/2}$ and $2p\, {}^2\!P_{\!3/2}$ excited levels. The only level that contributes to the scattering polarization and the Hanle effect of Ly$\alpha$ is the $2p\,{}^2\! P_{\!3/2}$ \citep[e.g.,][]{jtb-stepan-casini11}. We have adopted the same atomic data used by \citet{jtb-stepan-casini11} in their one-dimensional (1D) investigation, the only difference being that we have used a model atom composed of the above-mentioned levels and the $2s\,{}^2\! S_{\!1/2}$ level whose collisional coupling with the $P$-levels may contribute to a slight depolarization of the line. As shown by \citet{stepanjtb11a} this is a very suitable atomic model for computing the line-center polarization of Ly$\alpha$. We aim at comparing the $Q/I$ and $U/I$ profiles computed assuming $B=0$ G at each spatial grid point of the above-mentioned 2D atmospheric model with the fractional linear polarization signals calculated taking into account the Hanle effect caused by the model's magnetic field. To this end, we have applied the radiative transfer code mentioned in Section~1. As advanced by \citet{stepanjtb12a}, two novel features of this 3D code for the transfer of polarized radiation is a formal solver based on short-characteristics with quadratic Bezier-splines interpolation (which implies a significantly more stable formal solver than those based on the standard parabolic interpolation method) and a highly-convergent iterative method based on the non-linear multigrid method described in \citet{mgrid}. For the angular integration we have used Gaussian quadrature for the ray inclinations with respect to the vertical Z-axis (with 7 inclinations per octant) and the trapezoidal rule for the ray azimuth (with 5 azimuths per octant).

\begin{figure}[t]
\centering
\includegraphics[width=\columnwidth]{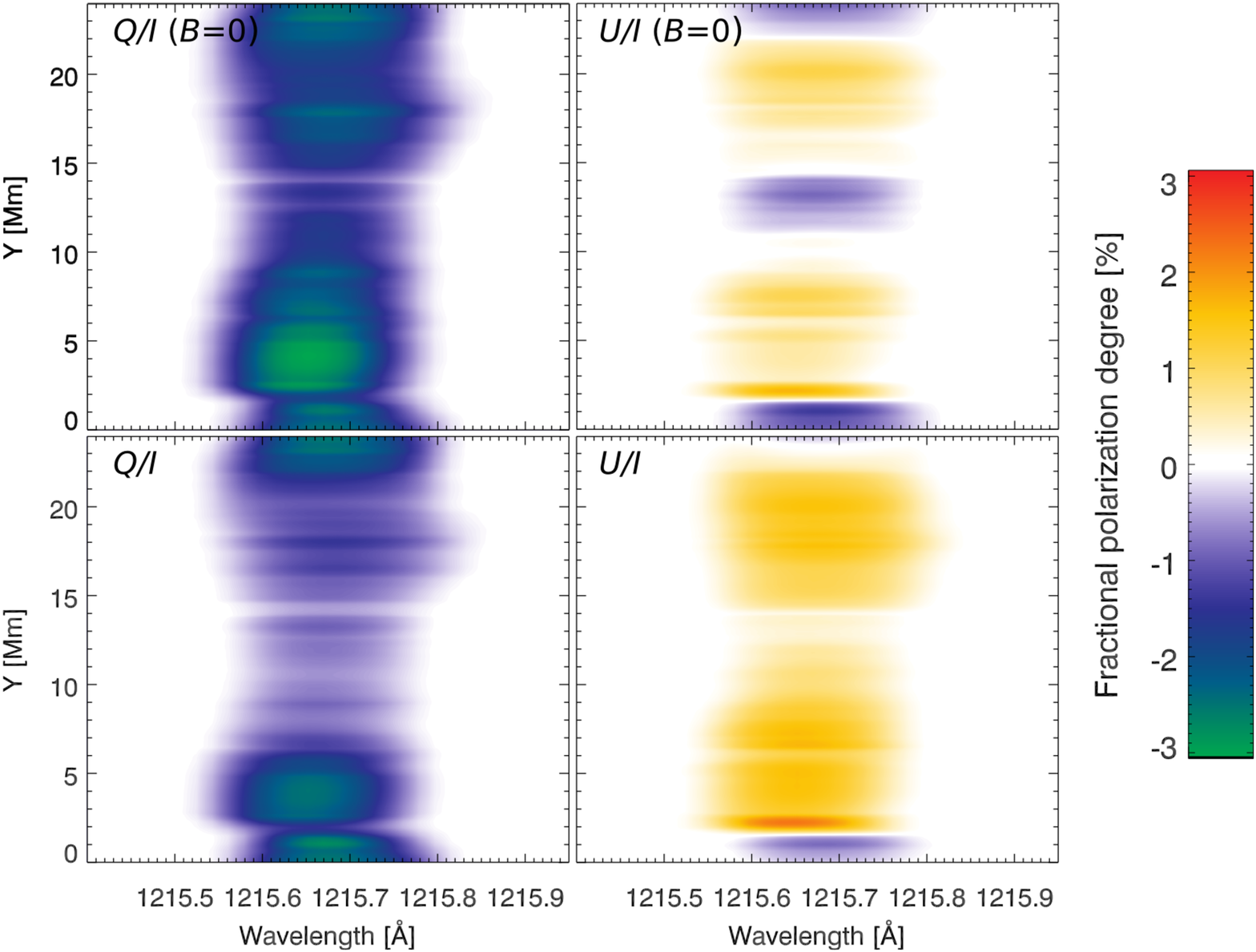}
\caption[]{The emergent $Q/I$ and $U/I$ profiles calculated at each Y-point in the 2D model of Fig.~\ref{fig:figure-2}, for a LOS with $\mu=0.3$ and $\chi=0^\circ$. Top panels: assuming $B=0$ G at each point within the 2D model. Bottom panels: taking into account the Hanle effect produced by the model's magnetic field. The Stokes profiles at any given Y position were averaged over a $5''$ interval.}
\label{fig:figure-3}
\end{figure}

The top panel of Fig.~\ref{fig:figure-2} shows the spatial variation of the kinetic temperature in the chosen 2D model atmosphere, where it can be clearly seen that the height of the transition region changes drastically from one horizontal location to another, outlining a highly crumpled surface. The next panel of Fig.~\ref{fig:figure-2} shows the spatial variation of the fractional anisotropy of the Ly$\alpha$ radiation, $\bar J^2_0/\bar J^0_0$ \citep[see equations (2) and (3) in][]{jtb-stepan-casini11}, calculated after obtaining with the above-mentioned radiative transfer code the self-consistent solution to the problem of resonance line polarization and the Hanle effect. As shown in this panel, the height variation of $\bar J^2_0/\bar J^0_0$, at most horizontal positions Y, is qualitatively similar to that shown for Ly$\alpha$ in figure~1 of \citet{jtb-stepan-casini11}, which corresponds to the solution of the same non-LTE polarization transfer problem in the semi-empirical model C of Fontenla et al. (1993; hereafter, FAL-C model). Note that at most horizontal positions the fractional anisotropy is zero at all photospheric and chromospheric heights, but that it suddenly becomes significant right at the location of the transition region, with larger negative values than in the case of the semi-empirical FAL-C model atmosphere. This is important because of the following two reasons: (1) the corrugated surface that delineates the model's transition region practically coincides with that corresponding to line-center optical depth unity along the line of sight (LOS), and (2) the line-center amplitude of the $Q/I$ profile of the Ly$\alpha$ line is approximately proportional to the $\bar J^2_0/\bar J^0_0$ value at such transition region heights \citep[cf., equation (1) of][]{jtb-stepan-casini11}. 

The extra three panels of Fig.~\ref{fig:figure-2} show the spatial variation of the strength, inclination and azimuth of the model's magnetic field. Note that at the crinkled surface that outlines the location of the model's transition region (which approximately coincides with the height where the Ly$\alpha$ line-center optical depth is unity along the LOS) the strength of the magnetic field varies between a few gauss and 60 gauss, approximately. Since the Hanle effect in Ly$\alpha$ is mainly sensitive to magnetic strengths between 10 and 100 G \citep[cf.,][]{jtb-stepan-casini11}, we should expect a significant impact on the emergent $Q/I$ and $U/I$ line-core signals.

\section{The emergent $Q/I$ and $U/I$ signals\label{sec:signals}}

The first step needed to compute the emergent $Q/I$ and $U/I$ signals is to obtain, at each spatial point within the chosen model atmosphere, the self-consistent values of the density-matrix elements corresponding to each $j$-level of our 4-level atomic model (i.e., 9 $\rho^K_Q$ elements quantifying the overall population of each $j$-level and the population imbalances and coherences among the sublevels of the $2p\,{}^2\!P_{\!3/2}$ level). To this end, we have applied the 3D radiative transfer code developed by \citet{stepanjtb12a,stepanjtb12b} for solving the two following cases: (a) the zero-field reference case, by forcing $B=0$ G at each spatial grid point in the atmospheric model (top panels) and (b) the Hanle effect case, by taking into account the model's magnetic field. The general multilevel expressions of the radiative transfer coefficients and of the statistical equilibrium equations corresponding to this so-called NLTE problem of the second kind can be found in Section 7.2 of \citet{landi-landolfi04}.

In Fig.~\ref{fig:figure-3} we show, for a LOS with $\mu=0.3$ and azimuth $\chi=0^{\circ}$, the emergent $Q/I$ and $U/I$ profiles calculated for each of the above-mentioned two cases. The first important point to emphasize is that the amplitudes of the linear polarization signals are very significant (${\sim}1\%$), both in the absence (top panels) and in the presence (bottom panels) of the model's magnetic field, and that they show significant spatial variations.

Note that $Q/I$ is always negative and that its amplitude is significantly larger in the zero field case. The fact that $U/I \ne 0$ for the $B=0$\,G case (top panels) indicates that the model's horizontal inhomogeneities (e.g., in temperature and density) are efficient in breaking the axial symmetry of the incident radiation field at each point within the medium. Interestingly enough, both the amplitude and the sign of $U/I$ are significantly modified by the action of the Hanle effect (compare the top and bottom panels), mainly in the central part of the 2D model where the magnetic field strength is significantly larger than in the model's boundaries (see Fig.~\ref{fig:figure-2}).

\section{The magnetic sensitivity of the spatially-averaged $Q/I$ and $U/I$ signals\label{sec:hanle}}

\begin{figure}[t]  
\centering
\includegraphics[width=\columnwidth]{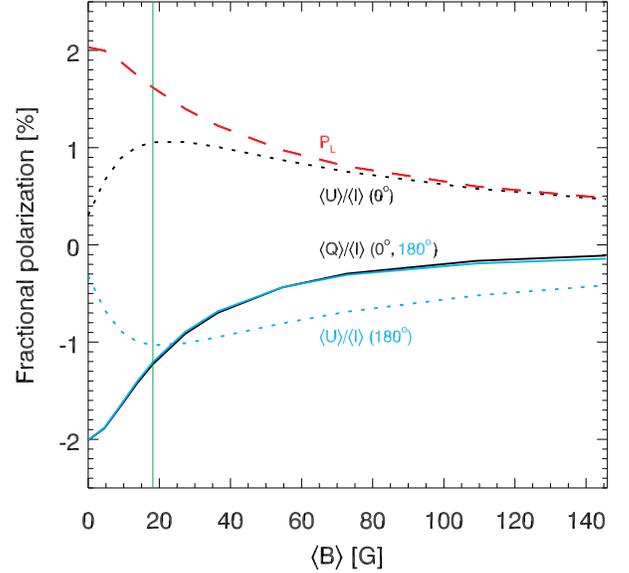}
\caption[]{The $\langle Q \rangle/\langle I \rangle$, $\langle U \rangle/\langle I \rangle$ and $P_{\rm L}=\sqrt{\langle Q \rangle^2+\langle U \rangle^2}/\langle I \rangle$ spatially-averaged amplitudes versus the mean field strength of the model's transition region, as discussed in Section 4. The LOS chosen have $\mu=0.3$ and azimuths $\chi=0^\circ$ (black lines) and $\chi=180^\circ$ (blue lines). The vertical green line indicates the mean field strength of the model's transition region. Other solutions have been obtained by scaling the modulus of the model's magnetic field, with a scaling factor $0\le\,f\le\,8$. Note that the $\langle Q \rangle/\langle I \rangle$ amplitude corresponding to the zero-field reference case in the chosen 2D model is significantly larger than that found by Trujillo Bueno et al. (2011) in the one-dimensional FAL-C model.}
\label{fig:figure-4}
\end{figure}

It is also interesting to consider the spatially-averaged $Q/I$ and $U/I$ signals, including their sensitivity to the mean magnetic field strength obtained by averaging the magnetic strength at each point of the corrugated surface that defines the spatial location of the model's transition region. To this end, we have multiplied the strength of the local magnetic field by a scaling factor $0\le\,f\le\,8$. Figure~\ref{fig:figure-4} shows the results for an LOS with $\mu=0.3$.

The dashed red line of Fig.~\ref{fig:figure-4} shows the magnetic sensitivity of the total fractional linear polarization, ${\rm P}_{\rm L}=\sqrt{\langle Q \rangle^2+\langle U \rangle^2}/\langle I \rangle$, where it can be seen that the model's magnetic field produces via the Hanle effect a significant depolarization with respect to the zero-field reference case. More interesting is to note that $\langle U \rangle/\langle I \rangle\approx 0$ in the absence of magnetic fields while $\langle U \rangle/\langle I \rangle\approx\pm 1\%$ with the model's magnetic field (see the black dotted lines). As shown by the blue solid lines, the Hanle effect of the model's magnetic field has a significant impact on the $\langle Q \rangle/\langle I \rangle$ amplitude, whose sign remains the same for the two considered LOS azimuths.

\section{Concluding comments\label{sec:conclusions}}

In order to obtain quantitative information on the magnetic field of the upper chromosphere and transition region of the Sun we need to measure and model the linear polarization caused by scattering processes in some strong resonance lines, such as the hydrogen Ly$\alpha$ line at 1216\,\AA. This is because via the Hanle effect such polarization is sensitive to the strength and orientation of the magnetic field expected for the solar transition region plasma (with $B\lesssim 100$\,G). The interpretation of the observed Stokes profiles will require taking into account that the outer solar atmosphere is highly structured and dynamic. For this reason it is important to do numerical simulations of the scattering polarization that is produced in realistic MHD models of the solar atmosphere, paying particular attention to understanding the impact of the model's magnetic field on the emergent $Q/I$ and $U/I$ profiles.

In this Letter we have made a first step towards that goal, by solving the multilevel radiative transfer problem of the scattering polarization and Hanle effect in Ly$\alpha$ in a 2D snapshot taken from the 3D radiation MHD simulation of an enhanced network region described in Leenaarts et al. (2012). To this end, we applied the 3D radiative transfer code described in \citet{stepanjtb12a,stepanjtb12b}.

Our model calculations of the line-core $Q/I$ and $U/I$ signals at each point on the model's surface show that the horizontal inhomogeneities of the chosen atmospheric model produce a significant impact on the local scattering polarization amplitudes, and that these linear polarization signals are significantly different in the absence and in the presence of the Hanle effect produced by the model's magnetic field. Moreover, the spatially-averaged $Q/I$ and $U/I$ signals show also an interesting sensitivity to the model's magnetic field. We therefore conclude that by measuring the polarization signals produced by scattering processes and the Hanle effect in Ly$\alpha$ and contrasting them with those computed in increasingly realistic atmospheric models, we should be able to decipher the magnetic, thermal and dynamic structure of the upper chromosphere and transition region of the Sun.

\acknowledgements
We are grateful to Rafael Manso Sainz (IAC) for carefully reviewing this Letter and for several interesting discussions. Financial support by the Grant Agency of the Czech Republic through grant \mbox{P209/12/P741} and project \mbox{RVO:67985815}, and by the Spanish Ministry of Economy and Competitiveness through projects \mbox{AYA2010--18029} (Solar Magnetism and Astrophysical Spectropolarimetry) and CONSOLIDER INGENIO CSD2009-00038 (Molecular Astrophysics: The Herschel and Alma Era) is gratefully acknowledged. This research was supported also by the Research Council of Norway through the project ``Solar Atmospheric Modelling'' and through grants of computing time from the Programme for Supercomputing. JL recognizes support from the Netherlands Organization for Scientific Research (NWO).

\end{document}